\documentclass[11pt, a4paper]{article}
\usepackage{moriond,epsfig}

\def\be{\begin{equation}}
\def\ee{\end{equation}}
\def\bea{\begin{eqnarray}}
\def\eea{\end{eqnarray}}

\begin{document}
\vspace*{4cm}
\title{Herwig++}

\author{Martyn Gigg and Peter Richardson}
\address{Institute of Particle Physics Phenomenology, Department of Physics, University of Durham, Durham, DH1 3LE, U.K..}

\maketitle\abstracts{We describe the recent development of the {\textsf Herwig++} event generator.}
\noindent
{\small¥{\it Keywords}: Hadron Colliders, Monte Carlo Simulations}

\section{Introduction}

  Monte Carlo event generators have become an essential part of all
  experimental analyses in particle physics. While it has been possible to extend and
  improve the existing HERWIG~\cite{HERWIG} program over many years making
  further improvements is increasingly difficult. 
  In order to include our improved understanding 
  of the physics and recent theoretical developments a programme is therefore 
  underway to 
  produce a new simulation in C++, 
  {\textsf Herwig++}~\cite{Gieseke:2003hm,Gieseke:2006ga},
  based on the same physics philosophy and models. This is part of a wider
  program within the Monte Carlo community to produce a new generation of 
  simulations for the LHC.~\cite{new}

  We will present the recent physics developments in the {\textsf Herwig++}
  simulation, concentrating on improvements to the simulation of QCD radiation and new
  physics, and plans for further improvements.

\section{Simulation of QCD Radiation}

  The main change between the HERWIG and {\textsf Herwig++}
  programs is in the simulation of 
  perturbative QCD radiation. While both programs use an angular-ordered parton
  shower designed to treat soft-gluon interference effects the algorithm 
  in {\textsf Herwig++}~\cite{Gieseke:2003hm,Gieseke:2003rz}
  has a number of improvements: invariance under boosts along the jet axis; improved
  treatment of radiation from heavy quarks; and better coverage of the soft region
  of phase space. 

  In particular the new algorithm uses an improved  evolution variable and
  the quasi-collinear splitting functions~\cite{Catani:2000ef} to give better
  treatment of the radiation from massive particles. In FORTRAN HERWIG the ``dead-cone''
  approximation of forbidding radiation with angle less than $m/E$, where $m$ is
  the mass and $E$ the energy of the heavy particle, was used. In {\textsf Herwig++} 
  this is replaced by a smooth suppression of radiation in the direction of the 
  particle. 

  The improved treatment of the kinematics of the branchings in the shower
  means that the soft-region of phase space in 
  $e^+e^-\to q\bar{q}$ is smoothly covered with 
  radiation from the quark and anti-quark filling separate regions of phase space
  which cover the whole region for soft emission without overlapping, as was the 
  case with the FORTRAN algorithm. 

  A major new feature is the inclusion of radiation from the decaying particle in heavy
  particle decays, for example $t\to bW^+$. This means that in these decays the 
  soft region for gluon emission 
  is fully covered, whereas in the FORTRAN program, which only included
  radiation from the decay products, part of the soft region was not 
  filled.~\cite{Corcella:1998rs} This makes correcting the parton shower
  using the exact single emission matrix element simpler. These corrections are now
  included for $e^+e^-\to q\bar{q}$, top decay~\cite{Hamilton:2006ms} 
  and the Drell-Yan process.

  Another key feature of the new algorithm is that momentum reshuffling needed to
  ensure energy and momentum conservation is under greater analytic
  control which will make it easier to match with higher order matrix elements. A
  number of developments in this area are underway.~\cite{Latunde-Dada:2006gx}

\section{BSM Physics}

\begin{figure}
\begin{center}
\includegraphics[angle=90,width=0.45\textwidth]{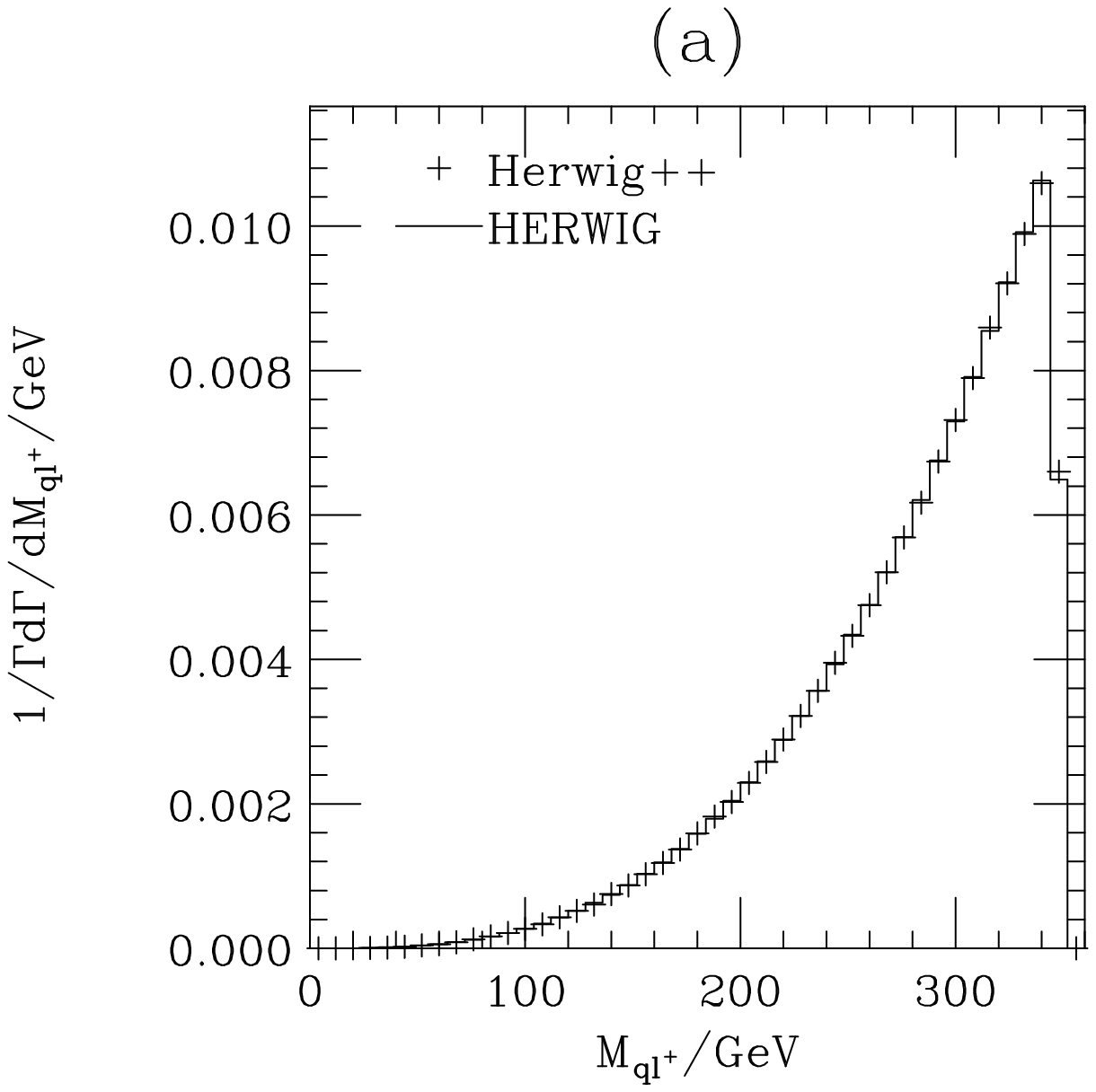}
\includegraphics[angle=90,width=0.45\textwidth]{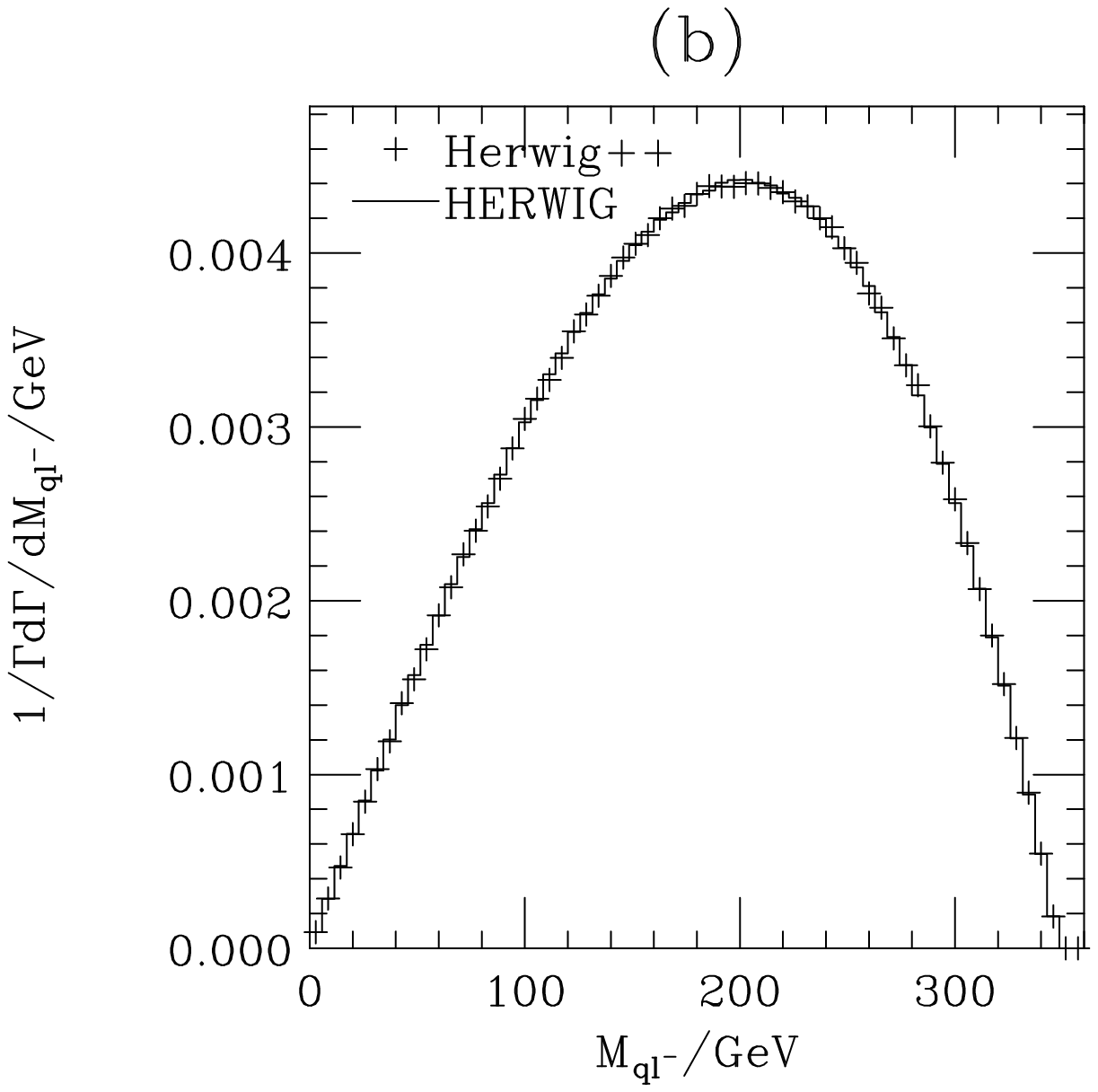}\\
\caption{Mass distribution of the quark and lepton in the decay 
	$\tilde{q}_L\to q\tilde{\chi}^0_2\to q\ell^\pm\tilde{\ell}_R^\mp$
	for (a) $\ell^+$ and (b) $\ell^-$.}
\label{fig:susyql}
\end{center}
\end{figure}

  The existing HERWIG program includes a detailed simulation of supersymmetric~(SUSY)
  models~\cite{Moretti:2002eu}
  including both spin correlation effects \cite{Richardson:2001df}
  and R-parity violating models.~\cite{Dreiner:1999qz}
  However, while the simulation of SUSY models was highly sophisticated, extending
  the simulation to other models of new physics was difficult and time consuming.
  In the new simulation we have adopted an entirely different approach
  for the inclusion of new physics models.~\cite{Gigg:2007cr} 
  In the FORTRAN simulation the matrix element for each new scattering process 
  and decay was added by hand. In the new simulation we have included a library 
  based on the HELAS \cite{Murayama:1992gi} formalism which is used in all matrix
  element calculations.
  The spin structures for the possible $2\to2$ matrix elements and $1\to2$ decays
  are included, based on the possible Feynman diagrams for each combination of the spins
  of external particles. The possible scattering processes and decays are then 
  automatically calculated from the Feynman rules implemented in the code. Using
  the HELAS formalism allows us to include spin correlations in the decays of the 
  fundamental particles, and also using the new simulation of tau and hadron decays
  the correlations in these decays which can be important in the decay of SUSY
  particles and the Higgs boson.
 
  This approach was originally tested using the Randall-Sundrum 
  model~\cite{Randall:1999ee} and the Minimal
  Supersymmetric Standard Model~(MSSM). An example of the mass distribution of the 
  quark and lepton produced in the decay
  $\tilde{q}_L\to q\tilde{\chi}^0_2\to q\ell^\pm\tilde{\ell}_R^\mp$, is shown
  in Fig.\,\ref{fig:susyql}. It is important that the correlations in this decay
  are correct as it may be possible to measure the spins of the SUSY particles
  using this decay mode.~\cite{Barr:2004ze}

\begin{figure}
\begin{center}
\includegraphics[angle=90,width=0.45\textwidth]{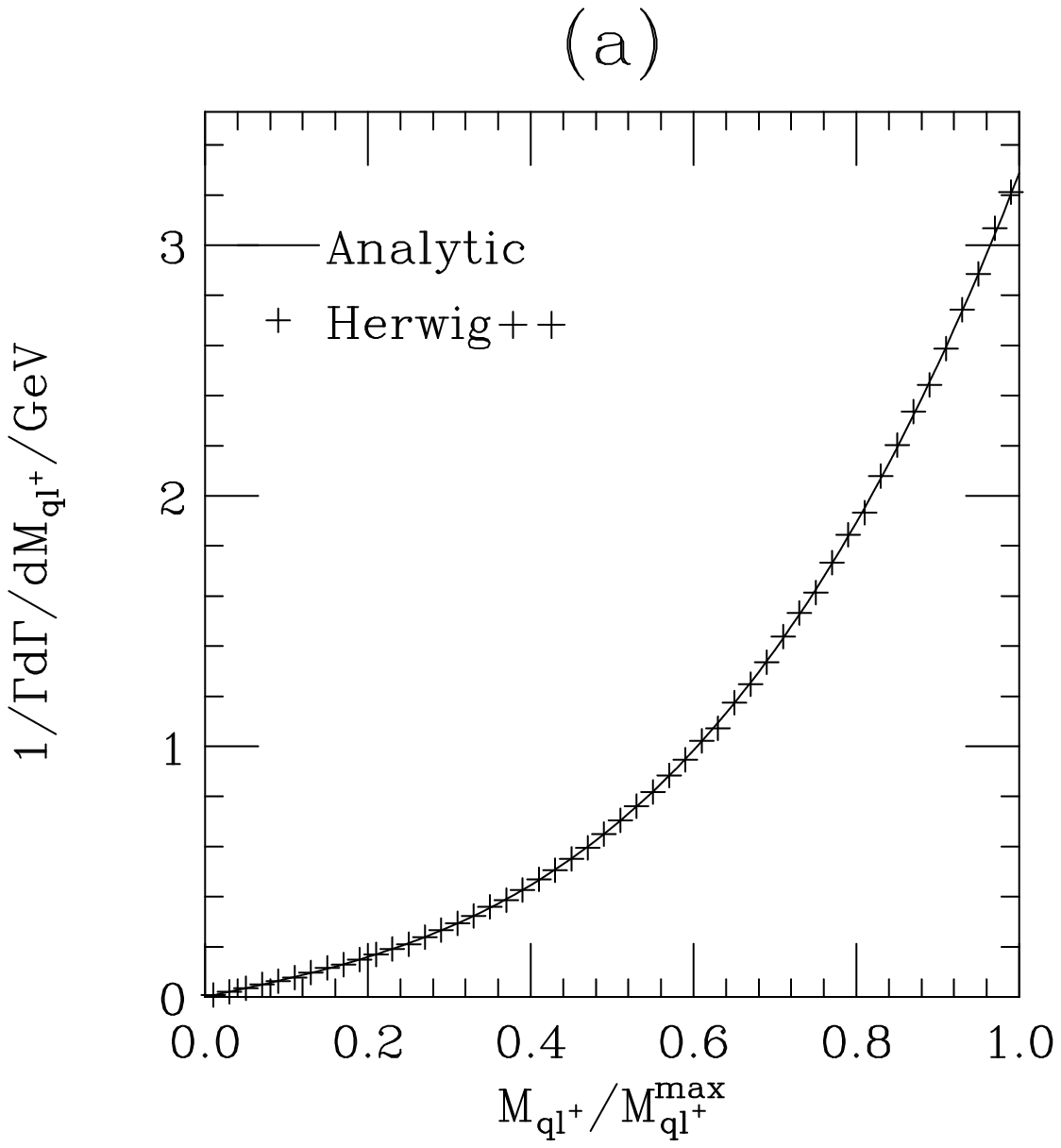}
\includegraphics[angle=90,width=0.45\textwidth]{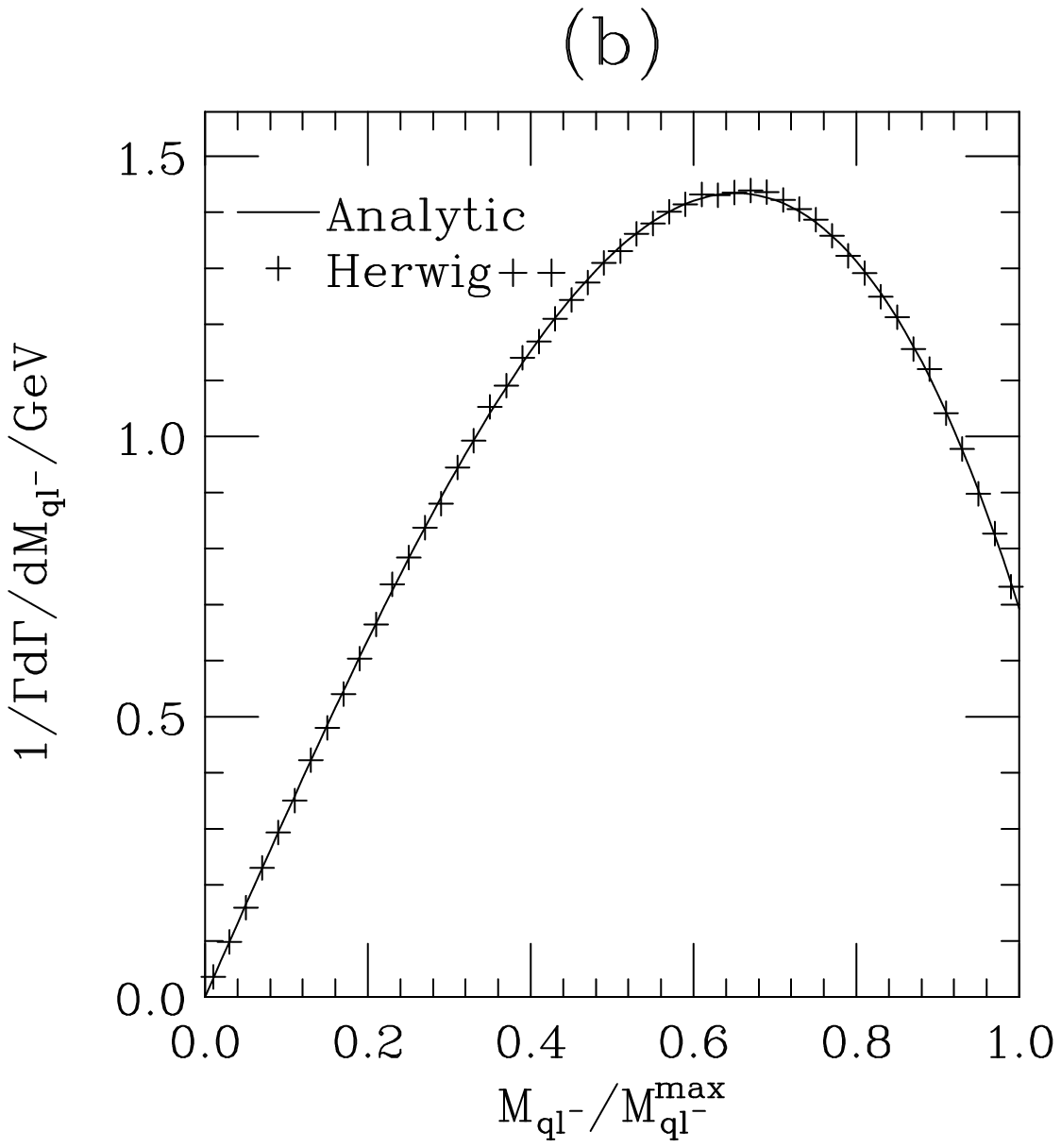}\\
\caption{Mass distribution of the quark and lepton in the decay
	$q^*_L\to q Z^* \to \ell^\pm \ell^{*\mp}_R$ in the UED model
	for (a) $\ell^+$ and (b) $\ell^-$. The results of {\textsf Herwig++} 
	are compared with the analytic results.~{\protect \cite{Smillie:2005ar}}
       The mass is given in terms of the maximum possible value.}
\label{fig:UEDql}
\end{center}
\end{figure}

  An important test of our new approach for the simulation of BSM physics is the 
  inclusion of additional models. We have therefore included the Universal Extra
  Dimensions model.~\cite{UED} This model is a useful ``straw-man'' as its particle
  content is similar to the MSSM but the new particles have the same spin as their
  Standard Model counterparts, rather than opposite spin statistics as in SUSY.
  The mass distribution of the quark and lepton pair in the equivalent decay chain to the SUSY
  chain shown in Fig.\,\ref{fig:susyql} is shown in Fig.\,\ref{fig:UEDql} and
  compared with the analytic results~\cite{Smillie:2005ar} for these distributions.

\section{Conclusions}

\begin{figure}
\begin{center}
\includegraphics[width=0.34\textwidth,angle=90]{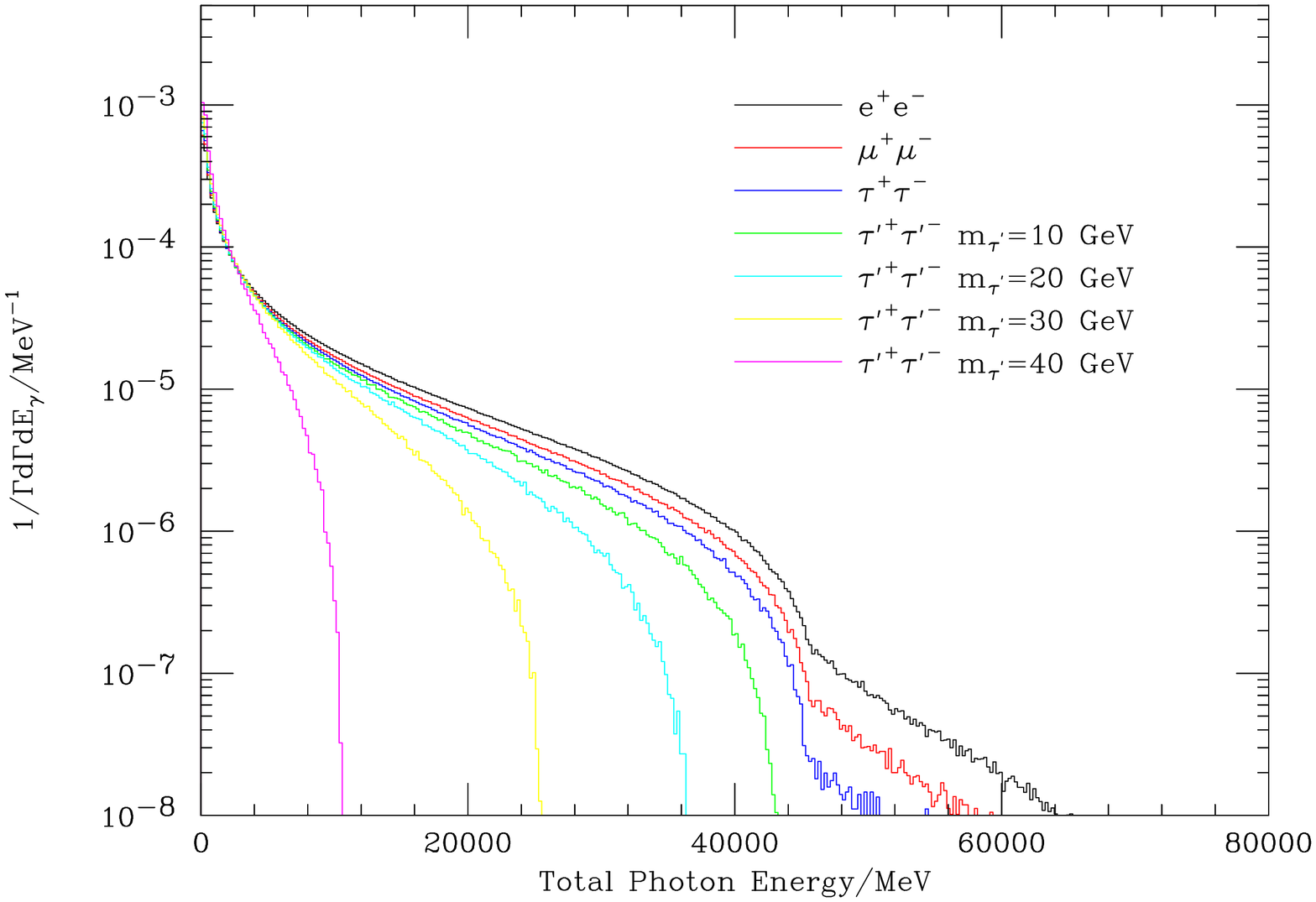}
\includegraphics[width=0.34\textwidth,angle=90]{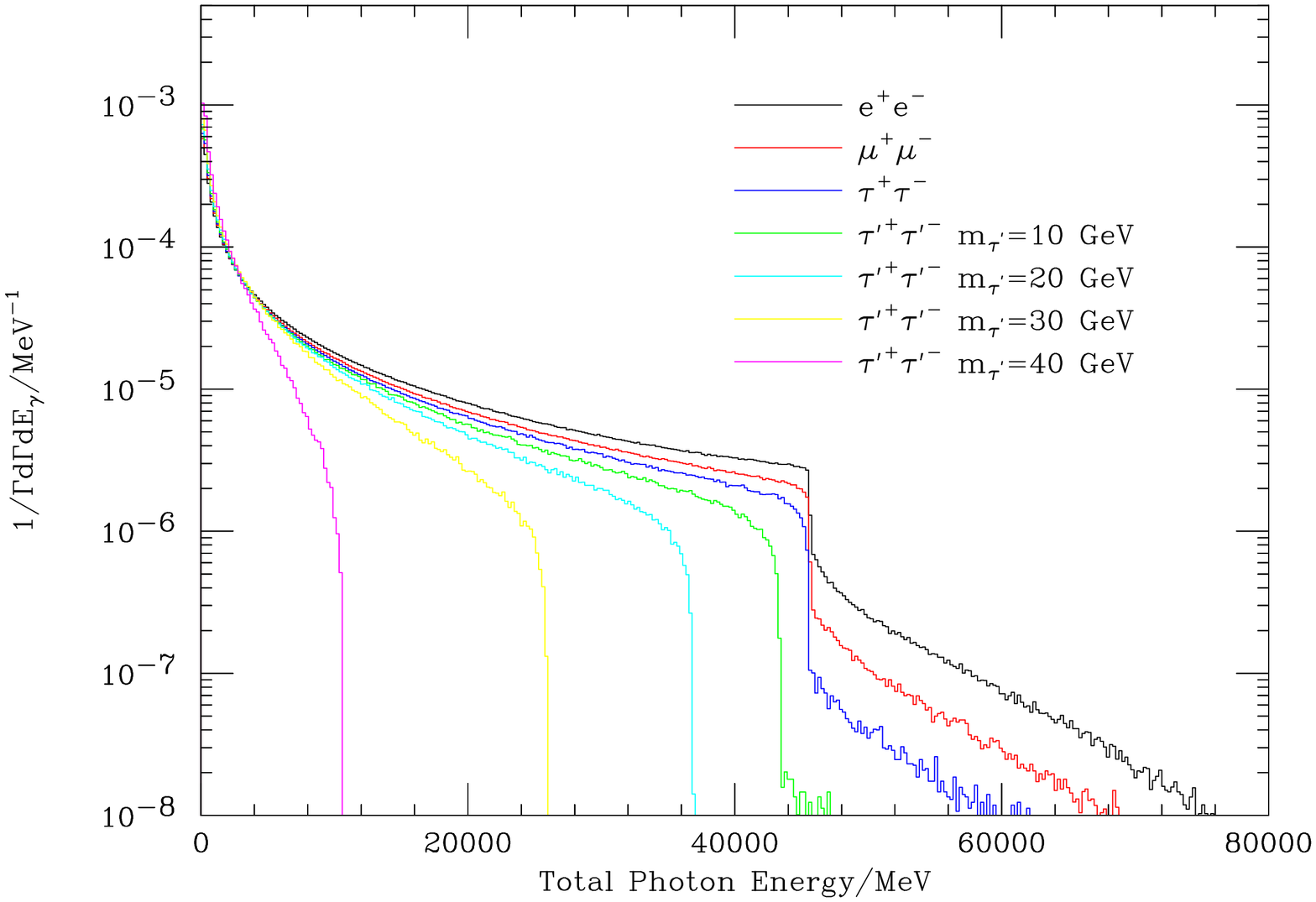}
\end{center}
\caption{
The total energy of the photons radiated in $\mathrm{Z}$ boson decays to leptons: 
(a) shows the spectrum when only soft radiation is included and
(b) shows the effect of including the collinear approximation for the hardest emission.}
\label{fig:YFS}
\end{figure}

  The {\textsf Herwig++} simulation 
  includes a number of improvements over the previous FORTRAN
  version for both the simulation of perturbative QCD radiation and simulation of 
  new physics.
  In addition to the improvements in the simulation of the perturbative stages of 
  the event generation process presented here improvements have been made to the 
  simulation of QED radiation~\cite{Hamilton:2006xz}~(an example of
  the radiation in Z decays is shown in Fig.\,\ref{fig:YFS}) and hadron and tau decays.
  
  The most recent version of {\textsf Herwig++}~\cite{Gieseke:2006ga} is now ready for
  the simulation of hadron collisions and further improvements will be available in
  the near future.

\section*{Acknowledgements}

  We would like to thank our collaborators on the {\textsf Herwig++} project
  upon whose work the material presented here is based. This work
  was supported in part by the Science and Technology Facilities Council and the 
  European Union Marie Curie Research Training Network MCnet under
  contract MRTN-CT-2006-035606.

\end{document}